\documentclass{article}
\usepackage{color}
\usepackage{amsmath}
\def\DH{\rm I\kern-1.5pt\rm H\kern-1.5pt\rm I}

\input epsf

\def\DR{\rm I\kern-1.45pt\rm R}
\def\DC{\kern2pt {\hbox{\sqi I}}\kern-4.2pt\rm C}

\newcommand{\ba}{\begin{array}}
\newcommand{\ea}{\end{array}}
\newcommand{\be}{\begin{equation}}
\newcommand{\ee}{\end{equation}}
\newcommand{\bea}{\begin{eqnarray}}
\newcommand{\eea}{\end{eqnarray}}
\newcommand{\bi}{\begin{itemize}}
\newcommand{\ei}{\end{itemize}}

\newcommand\ointint{\begingroup
\displaystyle \unitlength 1pt
\int\mkern-12.2mu
\begin{picture}(0,3)
\put(0,1){\oval(5,5)}
\end{picture}\mkern 6mu
\endgroup}

\begin{document}
\thispagestyle{empty}
\begin{center}
{\bf \Large Quantum  ring models and action-angle variables}\\
\vspace{0.5 cm} {\large  Stefano Bellucci${}^a$, Armen Nersessian${}^b$, Armen Saghatelian${}^b$ and
Vahagn Yeghikyan${}^b$}
\end{center}
{\sl${}^a$INFN-Laboratori Nazionali di Frascati, Via E. Fermi 40,
00044 Frascati, Italy}\\
{\sl ${}^b$Yerevan State University, 1 Alex Manoogian St.,   Yerevan,
0025, Armenia}
\begin{abstract}
We suggest to use the action-angle variables for the study of properties
of (quasi)particles in quantum rings.
 For this purpose we
present the action-angle variables for three two-dimensional singular oscillator systems.
 The first one is the usual (Euclidean) singular
oscillator, which plays the role of the confinement potential for the quantum ring.
 We also propose two singular spherical oscillator models for the role
 of the confinement system for the spherical ring.
 The first one is based on the standard Higgs oscillator potential.
 We show that, in spite of the
 presence of a hidden symmetry, it is not convenient for the study of the
 system's behaviour in a magnetic field.
 The second model is based on the
 so-called $CP^1$ oscillator potential and respects
 the inclusion of a constant magnetic field.

\end{abstract}


\section{Introduction}
It is  well-known that, for the systems with finite motion, one can
introduce the distinguished set of phase space variables (the
``action-angle" variables),   such that the ``angle" variables
parameterize the 
torus, while their conjugated ``action" variables are
the functions of constants of motions only \cite{arnold}. As a
consequence, the
 Hamiltonian depends only on action variables.
The formulation of the integrable system  in these variables gives us
a comprehensive  geometric description of its dynamics. Such a
formulation defines a useful tool for the developing of perturbation
theory, since the ``action" variables define adiabatic invariants of
the system.
 The action-angle formulation is important
from the quantum-mechanical point of view as well,
 since in action-angle variables the Bohr-Sommerfeld quantization
 is equivalent to the canonical quantization,
with trivial expressions for the wavefunctions.
 Hence,  evaluation of quantum-mechanical
aspects of such system becomes quite simple in this approach.

So, the action-angle variables form a useful tool for the study   of
systems with finite motion. But just such systems presently
attract much attention
 because of progress in mesoscopic physics, where
we usually deal with a motion of (quasi)particles localized in
quantum dots, quantum layers etc. On the other hand, due to the
recent progress in nanotechnology, now the fabrication of various
low-dimensional systems of complicated geometric form (nanotubes,
nanofibers, spherical and cylindrical layers) become possible
\cite{nano}. In these context the  methods of quantum mechanics on
curved space should be relevant for the description of the physics of
nanostructures. Say, the common method for the localizing of the
particle in the disc or in the cylinder is that of the
two-dimensional oscillator for  the role of the 
confinement potential.
Similarly, for the localization of
 the particle quantum lens (e.g. $GaAs/In_{1-x}Ga_xAs$, see \cite{lens})
  one can use the Higgs model of the spherical oscillator
  defined by the potential $V_{Higgs}=\frac 12 \omega^2r^2_0\tan^2\theta $  \cite{higgs}.
  Another confinement potential which could be used for the localization of the (quasi)particles
  in quantum lens, is the potential of the so-called $CP^1$ oscillator
  $V_{Higgs}=2 \omega^2r^2_0\tan^2\theta/2 $ \cite{Bellucci}.
The advantage  of the latter potential is with
   respect to the  magnetic field, which has a constant magnitude on the surface of the sphere.
   Such a magnetic field is precisely the magnetic field of a Dirac monopole located at the center of sphere.
   So, formally this field is an unphysical one. However, due to the restriction of the electron in the
   segment/ring of the spherical layer, it could be viewed as a physical field generated e.g. by the pole of
   a magnetic dipole.
The fabrication of semiconductor ring-shaped systems \cite{qr},
presently referred to as  quantum rings (e.g.
$In(Ga)As$ - two-dimensional quantum rings), led to the use of the
singular oscillator potential with the role of the confinement one.
A pioneering  work on the  theoretical study of the
impact of the magnetic field on the  electron properties in a quantum
ring was  written by Chakraborty and Pietelainen \cite{qdch}.
 There, for the role of the confinement potential
restricting the motion of electrons in the quantum ring the
shifted oscillator potential $V_{\rm ChP}=\beta(r-r_0)^2$ was choosen. The
results obtained within this approximation are in
 a good correspondence with experimental data. The
quantum ring model of Chakraborty  and Pietelainen  is
not exactly solvable in the general case, and it assumes
 the use of numerical
 simulations.
 The quantum ring model based on the singular oscillator system  \cite{qdsw} has been suggested
  as an analytically  solvable alternative to the Chakraborty-Pietelainen model.
 Although calculations performed within the Chakraborty-Pietelainen model are
 in  better correspondence with experimental data
 than those within the singular oscillator potential \cite{comp}, the latter has its own place
 in the study of quantum rings (see,e.g. \cite{hayk}).

In analogy with the above models, one can suppose, that
 singular versions of two-dimensional Higgs and $CP^1$ oscillators may
be appropriate candidates for the confinement
 potential localizing the motion of the electron in the ring of a spherical quantum layer.

By the above listed reasons  we present, in this paper, the action-angle
formulation of the two-dimensional singular oscillator and of its
spherical generalizations based on Higgs and $CP^1$ spherical
oscillator models.

We shall start from the  simple model of a two-dimensional singular
oscillator given by the Hamiltonian \be H=\frac{{\bf
p}^2}{2}+\frac{\alpha^2}{2{\bf r}^2} + \frac{\omega^2 {\bf r}^2}{2}.
\label{singosc} \ee Then we shall consider a two-dimensional singular
spherical oscillator defined  by the following Hamiltonian: \be
H_{\rm Higgs}=\frac{p_{\theta}^2}{2r^2_0}+\frac{
p_{\varphi}^2}{2r^2_0 \sin ^2\theta}+\frac{\alpha^2}{2r^2_0}
\cot^2\theta+\frac{\omega^2r^2_0}{2} \tan^2\theta ,\label{higgs}\ee
where $r_0$ is the radius of the sphere.

This system generalizes the well-known Higgs model of the spherical
oscillator \cite{higgs}, whose
uniqueness is in the closeness
  of all trajectories,
 which reflects   the existence of a number of  hidden symmetries    equal to the those
 of the Euclidean oscillator.
By this reason, the Higgs oscillator is a convenient background for the
developing of perturbation theory. Particularly, it admits the
anisotropic modification  preserving the integrability of the system
\cite{anosc}. Hence, such a model of the spherical ring should be
convenient for the study of electrons behavior in external
potential fields, e.g., in the electric one. However, it is easy to
observe that the (singular) Higgs oscillator does not preserve its
exact solvability in the presence of a constant magnetic field, in
contrast with the Euclidean one. While the study of quantum dot
systems in a magnetic field is of a special physical importance.
By this reason we consider  the  alternative model of the singular
spherical oscillator, given by the Hamiltonian \cite{aramyan} \be
H_{CP^1}= \frac{p_{\theta}^2}{2r^2_0}+\frac{ p_{\varphi}^2}{2r^2_0
\sin
^2\theta}+\frac{\alpha^2}{8r^2_0}\cot^2\frac{\theta}{2}+2\omega^2r^2_0\tan^2\frac{\theta}{2}.
\label{alt}\ee It is based on the model of the 
oscillator on complex
projective spaces \cite{Bellucci} and, in contrast with the (singular)
Higgs oscillator, it respects the inclusion of a constant magnetic
field (of the Dirac monopole). Let us notice that a similar model on the
four-dimensional sphere and hyperboloid respects the inclusion of
the BPST instanton field \cite{mn}. Quantum mechanical solutions of
(\ref{alt}) are not constructed yet. But they could be found by
a proper modification of the solutions of the corresponding non-singular
system (third  reference in \cite{Bellucci}). Because of the absence of
hidden symmetries, this model is not convenient for the study of the
system in external potential (e.g. electric ) fields. But it
convenient for the study of the interaction with the external magnetic
field.

\setcounter{equation}{0}
\section{Action-angle variables}
The well-known Liouville theorem gives the exact criterium of integrability of the
$N$-dimensional mechanical system: that is the existence of   $N$ mutually commuting constants
of motion $F_1=H,\ldots, F_n$: $\{F_i, F_j\}=0$,$i,j=1,\ldots N$.
The theorem also states that if the level surface $M_f=\left( (p_i,q_i):
F_i=const\right)$ is a {\sl compact and connective manifold}, then it
is diffeomorphic to the $N$-dimensional torus $T^N$. The natural angular
coordinates ${\bf \Phi}=(\Phi_1,\ldots, \Phi_N)$ parameterizing that torus satisfy the motion
equations of a free particle moving on a circle. These coordinates form,  with
their conjugate momenta ${\bf I}=(I_1,\ldots, I_N)$, a full set of phase space  variables
called ``action-angle'' variables. One of the
results of the theorem is that the momenta ${\bf I}$ depend on constants of
motion only ${\bf I}={\bf I}({\bf F})$. So, there exists a
canonical transformation to the new variables $({\bf p},{\bf q})\mapsto
({\bf I},{\bf \Phi})$, in which the Hamiltonian depends  on the constants of
motion ${\bf I}$ (which are called action variables) only. Consequently, the
equations of motion read
 \be
\frac{d{\bf I}}{dt}=0,\quad \frac{d{\bf \Phi}}{dt}=\frac{\partial
H(I)}{\partial {\bf I}}
\qquad \{ I_i, \Phi_j
\}=\delta_{ij},\qquad  \Phi_i\in [0,2\pi ),\quad i,j=1,\ldots, N.
 \ee
Besides the practical importance, the action-angle formulation has an academic interest as well.
From the academic viewpoint, it gives a 
precise indication of the (non)equivalence of different
Hamiltonian systems. Indeed, gauging the integrable system by
action-angle variables, we preserve the freedom only in the
functional dependence of the Hamiltonian from the action
variables, $H=H({\bf I})$, and
in the range of validity of the action variables, $I_i\in [\beta^{-}_i, \beta ^{+}_i]$.
Hence formulating the systems in terms of action-angle variables, we can indicate the (non)equivalence
of different integrable systems. Let us refer, in this respect, to the recent paper \cite{lny}, where, particularly,
the global equivalence of $A_2$ and $G_2$ rational Calogero models, and their global equivalence
with a free particle on the circle, has been established in this way.

In action-angle variables the Bohr-Sommerfeld quantization is equivalent to the canonical quantization,
with a quite simple  expression for the wavefunction
\be {\widehat I}_i\Psi(\Phi )= I_i\Psi(\Phi),\qquad
{\widehat I}_i=-\imath\hbar \frac{\partial}{\partial \Phi_i},\quad
\Psi=\frac{1}{({2\pi})^{N/2}}\prod_{i=1}^{N}{\rm e}^{-\imath\; n_{i}\Phi_i},\quad I_i=\hbar  n_{i},
\label{spectra}\ee
where $n_{i}$ are integer numbers taking their values at the range
$[\beta^{-}_i, \beta^{+}_{i}]$.

The general prescription for the construction of action-angle
variables  looks  as follows \cite{arnold}.
 In order to construct the action-angle  variables, we should fix the level
surface of the Hamiltonian
 ${\bf F}={\bf c}$ and then introduce the generating function
 for the canonical transformation $({\bf p},{\bf q})\mapsto({\bf  I},{\bf \Phi})$,
which is defined by the expression
\be S({\bf c}, {\bf q})=\int_{{\bf F}={\bf c}} {\bf p} d{\bf q},\label{sdef} \ee where ${\bf p}$
are expressed via ${\bf c}, {\bf q}$ by the use of the constants of
motion.
The action variables ${\bf I}$ can be obtained from the expression \be
I_i(c)=\frac{1}{2\pi}\oint_{\gamma_i} {\bf p}d{\bf q}, \ee
where $\gamma_i$
is some loop of the level surface ${\bf F}={\bf c}$. Then inverting these relations, we can get the
expressions of ${\bf c}$ via action variables: ${\bf c}={\bf c}({\bf I})$.
The angle variables ${\bf \Phi}$ can be found from the expression \be
{\bf \Phi}= \frac{\partial S({\bf c}({\bf I}), {\bf q})}{\partial {\bf I}}. \ee
In the next sections we will use the above formulae for the construction of the ``action-angle" variables of the
two-dimensional singular oscillator models.

\setcounter{equation}{0}
\section{Singular Euclidean oscillator}
Let us demonstrate our approach with the simplest example of the
singular oscillator on the two-dimensional Euclidean space, which is defined by the
Hamiltonian (\ref{singosc}).
In polar coordinates this Hamiltonian reads
\be H=\frac{p_r^2}{2}+\frac{p_\varphi^2+\alpha^2}{2 r^2} +
\frac{\omega^2 r^2}{2},\qquad  x=r\cos\varphi,\quad y=r\sin\varphi.\label{2osc} \ee
Taking into account that the angular momentum $p_\varphi$ is the constant of motion of this system,
we can represent  its generating function as follows:
$S(p_\varphi, h,\varphi, r )= p_\varphi\varphi +\int_{H=h} p_r dr$.
So,
for the action variables we get the expressions
\be
I_1= \frac{1}{2 \pi} \ointint p_\varphi d \varphi = p_\varphi,\qquad
I_2=\frac{1}{2 \pi} \ointint p_r d r = \frac{h}{2\omega} - \frac{\widetilde{p}_\varphi}{2}
\qquad{\rm where}\quad\widetilde{p}_\varphi
\equiv\sqrt{p_{\varphi}^2+\alpha^2}
\ee
Respectively, the Hamiltonian takes the form
\be
H_{2d}=\omega\left(2I_2+\sqrt{I^2_1+\alpha^2}\right)
\ee The angle variables read
\be \Phi_1 = \varphi
-\frac{p_\varphi}{2\tilde{p}_\varphi}
\arcsin
\frac{\left(\widetilde{p}_{\varphi} + \omega r^2 \right) \sqrt{2 h r^2-\widetilde{p}_{\varphi}^2-\omega^2 r^4}}{(h+\widetilde{p}_{\varphi} \omega) r^2}
,\quad
\Phi_2 = 
-\arcsin\frac{h- r^2 \omega ^2}{\sqrt{h^2-\widetilde{p}_\varphi^2
\omega ^2}}.
\ee
For the reduction of this system to a one-dimensional one, we should put $p_\varphi=0$. In that case the
Hamiltonian takes the form (where we replaced  $r$ by $x$)
$H_{1d}=\omega(2I_2+{\alpha})\equiv 2\omega\tilde{I},\quad \tilde{I} \in [{\alpha}/{2},\infty)$.
So, in the action-angle variable the one-dimensional singular oscillator is locally
equivalent to the nonsingular one. The only difference is in the range of validity of the action variable.
Hence, in spite of the close similarity in the action-angle variable formulation of the one- and
two-dimensional singular
oscillators, the dependence on the singularity term in the second system cannot
be removed by the change in the range of validity of the action variable $\tilde{I}$,
in contrast with the one-dimensional case.
Let us notice that the action variable corresponding to the cyclic coordinate $\varphi$ coincides with the
angular momentum
$I_1=p_\varphi$. However, the respective angle variable $\Phi_1$ is different from the initial angle $\phi$.
In other words, the ``radial" motion, encoded in the dynamic of $I_2$ and $\Phi_2$ variables, has an essential
impact
on the `` angular" motion. While the impact of $\varphi, p_\varphi $ variables in the radial motion is the
shift $\alpha^2\to\alpha^2+p_{\varphi}^2$.

The inclusion of the constant magnetic field in the two-dimensional oscillator system does not
essentially change its properties.
Indeed, it is defined, in the two-dimensional planar system, by the potential
\be {\cal A}=\frac{B_0}{2} (xdy-ydx) = \frac{B_0
r^2}{2} d\varphi \ee
Hence, including the constant magnetic field in the two-dimensional singular oscillator, we shall get
\be H=\frac{p_r^2}{2}+\frac{(p_\varphi-\frac{B_0 r^2}{2})^2}{2
r^2}+\frac{\alpha^2}{2r^2} + \frac{\omega^2 r^2}{2} \qquad \Leftrightarrow \qquad \widetilde{H}=
\frac{p_r^2}{2}+\frac{\widetilde{p}^2_\varphi}{2
r^2}+ \frac{\widetilde{\omega}^2 r^2}{2}, \label{2osc} \ee
where we use the notation
\be {\widetilde{p}_{\varphi}^2} =
{p_{\varphi}^2} + \alpha^2 ,\quad\widetilde{\omega}^2 =
\omega^2+\frac{B_0^2}{4},\quad
 \widetilde{H} =H + \frac{B_0 p_\varphi}{2}  \label{substit}\ee
Hence, we get the  Hamiltonian (\ref{2osc}).
Thus, the impact of the magnetic field  in the generating function $S(h,p_\varphi, r, \varphi)$ consists in the
replacement (\ref{substit}).
Respectively,
the action variables and Hamiltonian are defined by the expressions
\be
I_1=p_\varphi,\quad
I_2=
\frac{\widetilde{h}}{2\widetilde{\omega}} -
\frac{\widetilde{p}_\varphi}{2} \qquad\Rightarrow\qquad
H=\sqrt{\omega^2+(B_0/2)^2} \left(2I_2+\sqrt{I_1^2+\alpha^2} \right)-\frac{B_0I_1}{2}.
\label{hsaa}\ee
The explicit expressions for angle variables reads
\be
\Phi_1 =\varphi - \frac{p_\varphi}{2 \widetilde{p}_{\varphi}}
\arcsin
\frac{\left(\widetilde{p}_{\varphi} + \widetilde{\omega} r^2 \right) \sqrt{2 \tilde{h} r^2-\widetilde{p}_{\varphi}^2-\widetilde{\omega}^2 r^4}}{(\tilde{h} + \widetilde{p}_{\varphi} \widetilde{\omega}) r^2}
,\qquad
%
\Phi_2 =
- \arcsin\frac{\widetilde{h}- r^2
\widetilde{\omega}^2}{\sqrt{\widetilde{h}^2-\widetilde{p}_\varphi^2
\widetilde{\omega}^2}}.
\ee
It is seen that the magnetic field yields in the Hamiltonian the term linear on $I_1$,
in addition to the predictable change of the effective frequency $\omega\to\sqrt{\omega^2+B^2_0/4}$.

So, we constructed the action-angle variables for the
two-dimensional singular oscillator in the constant magnetic field.
In the next sections we shall consider a similar formulation for the models of singular spherical oscillators.

\section{Singular Higgs  oscillator}
In this Section we present the action-angle formulation of the singular Higgs oscillator (\ref{higgs}).
In our  consideration we assume the
unit radius of the sphere, $r_0=1$. The restoration of the
the arbitrary radius can be carried out by the obvious redefinition of the
Hamiltonian and the constants $\alpha,\omega$.

Since the angular momentum $p_\varphi $ is a constant of motion of the system, the
the generating function of the action-angle variables takes the form
\be
S=p_\varphi \varphi+\int p_\theta( h, p_\varphi, \theta ) d\theta ,
\ee
where $H_{\rm Higgs}=h$.
From this generating function we get the action variables
\be
I_1= \frac{1}{2 \pi} \ointint p_\varphi d \varphi = p_\varphi\;,
\qquad I_2=\frac{1}{2\pi}\ointint p_\theta d\theta =
\frac{1}{\pi}\int_{\theta_-}^{\theta_+} \sqrt{2\left(h-\frac{p_{\varphi }^2}{2 \sin
^2\theta }-\frac{\alpha^2}{2} \cot^2\theta-\frac{\omega^2}{2} \tan^2\theta\right)}
d\theta ,\label{11}\ee
where the integration limits $\theta_\pm $ are defined by the condition
\be
2h=\frac{p_{\varphi }^2}{ \sin
^2\theta_\pm }+{\alpha^2} \cot^2\theta_\pm+{\omega^2} \tan^2\theta_\pm .
\ee
To calculate the integral in the second expression, we introduce the notation
\be
\xi=\frac{{1}}{a}\left[\cos{2\theta}
+{b}\right],\qquad
a=\sqrt{1-2\frac{p_\varphi^2+\alpha^2+\omega^2}{2h+\alpha^2+\omega^2}+
\left(\frac{p_\varphi^2+\alpha^2-\omega^2}{2h+\alpha^2+\omega^2}\right)^2},
\quad b=-\frac{p_\varphi^2+\alpha^2-\omega^2}{2h+\alpha^2+\omega^2}.
\label{definitions}
\ee
In this terms the second integral in (\ref{11}) reads (its value can be found by the use of standard methods,
 see, e.g.\cite{fiht,lny})
\be I_2 = \frac{a^2 \sqrt{2h+\alpha ^2+\omega ^2}}{2\pi}
\int\limits_{-1}^1\frac{\sqrt{1-\xi^2}}{1-\left(a\xi+b\right)^2}d\xi
=  \frac{1}{2}\left(\sqrt{2h+\alpha ^2+\omega ^2}-
\sqrt{p_{\varphi }^2+\alpha ^2}- \omega \right).  \label{h1}\ee
Hence, the functional dependence of the Hamiltonian from the action variables is given by the expression
\be
H=\frac12\left(2I_2+\sqrt{I_1^2+\alpha^2}+\omega\right)^2-\frac{\alpha^2+\omega^2}{2}.
\label{h2}\ee
For $\Phi_1$ and $\Phi_2$ we get

\be \Phi_1 =
\varphi-\frac{p_\varphi}{\widetilde{p}_\varphi}\arcsin \xi+
\frac{p_\varphi}{\widetilde{p}_\varphi}\arctan\frac{1}{2\widetilde{p}_\varphi}\left[
\sqrt{\frac{(2h-p_{\varphi}^{2})^2-4
\omega^2\widetilde{p}_{\varphi}^2}{2 h+\alpha ^2+\omega ^2}} -
\frac{2 h+2\alpha^2+p_{\varphi}^2}{\sqrt{2 h+\alpha ^2+\omega ^2}}
\frac{1+\sqrt{1-\xi^2} }{2 \xi} \right]\label{h3} \ee
\be \Phi_2= -2 \arcsin
\xi
\label{h4}\ee
Here, as in the previous Section, we use the notation
\be
\widetilde{p}_\varphi=\sqrt{p_\varphi^2+ \alpha^2}
\ee

We presented the action-angle formulation of the singular Higgs oscillator (\ref{higgs}) on the sphere of unit
 radius $r_0=1$.
 The action-angle formulation of the system on the sphere  with  arbitrary value of $r_0$
 could be easily found from (\ref{h1})-(\ref{h4}) by  the replacement
 \be
 { H}_{r_0}= \frac{H}{r_{0}^2},\quad {\rm with }\quad {\omega}\to\omega r_{0}^2 .
\label{rep} \ee
 In that case the Hamiltonian (\ref{higgs}) is defined, in the action-angle variables,
 by the following expression:
\be
H=\frac{1}{2r^2_0}\left(2I_2+\sqrt{I_1^2+\alpha^2}+
\omega r^2_0\right)^2-\frac{\alpha^2}{2r^{2}_0}-\frac{\omega^2r^2_0}{2}.
\label{h3}
\ee
It is seen that, in the planar limit  $r_0\to\infty$,
it results in the Hamiltonian of the Euclidean singular oscillator (\ref{hsaa}) with $B_0=0$
(i.e. in the absence of constant magnetic field).
However, the singular Higgs oscillator does not respect the inclusion of constant magnetic field,
 in contrast with the Euclidean one.

Indeed, the magnetic field which has a constant magnitude on the sphere,
is the field of a Dirac monopole located at the center of sphere.
It is defined by the following one-form:
\be
A_D=s(1-\cos\theta)d\varphi,\qquad s=B_0r^2_0.
\ee
Hence, the Hamiltonian of the singular  Higgs oscillator
interacting with a constant magnetic field, is defined by the expression
\be H=\frac{p_{\theta}^2}{2r^2_0}+\frac{\left[p_{\varphi }-s \left( 1- \cos
\theta \right) \right]^2}{2r^2_0\sin ^2\theta }+\frac{\alpha^2}{2r^2_0} \cot^2\theta+
\frac{\omega^2r^2_0}{2} \tan^2\theta .\ee
Writing down the corresponding generating function we shall see
 that the impact of the magnetic field cannot be absorbed
by the proper redefinition of constants. Hence, the inclusion of the magnetic field breaks
the exact solvability of the (singular)  Higgs oscillator, so that the presented model is not
suitable for the study of the
 properties of spherical bands and length  in the external magnetic field.
 However, this models is relevant for the consideration of their properties in the external potential,
  e.g. the electric field.
 Moreover, one can further modify the Higgs oscillator potential providing it by the
 anisotropy properties preserving the
 integrability of the system \cite{anosc}. Such a system would be useful to consider
 the quantum dots model restricted
 from the sphere to the spherical segment.

\setcounter{equation}{0}
\section{Singular ${C}P^1 $ oscillator}
There is another  model of the spherical oscillator,
which  was introduced in \cite{Bellucci} as a proper generalization of the oscillator system
to the complex projective spaces.
It was further generalized for the quaternionic projective spaces, as well \cite{mn}.
Its
specific property was the respect of the constant magnetic fields.
Since the complex projective plane is  equivalent to the two- dimensional sphere, we can use
this model for
the definition of the two-dimensional magnetic oscillator.
In spherical coordinates the potential of this alternative model
of the spherical oscillator reads
\be
V_{CP^1}=2\omega^2r^2_0\tan^2\frac{\theta}{2}.
\ee
In contrast with yje Higgs oscillator,
it preserves the exact solvability property upon inclusion of the constant
 magnetic field.
Respectively, its singular version,  defined by the Hamiltonian (\ref{alt}) also remains
exactly solvable in the presence of a constant magnetic field, at least, classically \cite{aramyan}.
Quantum mechanical solutions of the $CP^1$  singular oscillator are not constructed yet.
But they could be found by a proper modification of the solutions of the corresponding
non-singular system \cite{Bellucci}.

Inclusion of the constant magnetic field yields the following
modification of the Hamiltonian (\ref{alt}): \be H=\frac{p_{\theta
}^2}{2r^2_0}+\frac{\left[p_{\varphi }-s \left( 1- \cos \theta
\right) \right]^2}{2r^2_0\sin ^2\theta }+2 \omega^2r^2_0
\tan^2\frac{\theta }{2}+\frac{\alpha^2}{8r^2_0}
\cot^2\frac{\theta}{2},\qquad s=B_0r^2_0 .\label{alt_magn}\ee As
before, we put, without loss of generality, $r_0=1$. The way of the
restoring of $r_0$ is obvious. Then, in a completely similar way as in the
previous cases, we can construct the action-angle variables of this
system. For the action variables $I_1$ and $I_2$ we get \be I_1=
p_\varphi,\quad I_2=\frac{1}{\pi}\int_{\theta_-}^{\theta^+}d\theta
\sqrt{2h-\frac{\left[p_{\varphi }-s \left( 1- \cos \theta \right)
\right]^2}{\sin ^2\theta }-4 \omega^2 \tan^2\frac{\theta
}{2}-\frac{\alpha^2}{4} \cot^2\frac{\theta}{2}}  \ee where
$\theta_\pm$ are defined by the equation \be
h=\frac{\left[p_{\varphi }+s \left( 1- \cos \theta_\pm \right)
\right]^2}{2\sin ^2\theta_\pm }+2 \omega^2
\tan^2\frac{\theta_\pm}{2}+ \frac{\alpha^2}{8}
\cot^2\frac{\theta_\pm}{2}. \ee The explicit expression for the
second integral looks as follows:
$$
 I_2 =\frac{a^2\sqrt{2h+4\omega ^2+\frac{\alpha^2}{4}+s^2}}{\pi  }
\int\limits_{-1}^1\frac{\sqrt{1-\xi^2}}{1-\left(a\xi+b\right)^2}d\xi
=
$$
\be
 =\sqrt{2 h+ s^2+\frac{\alpha^2}{4}+4 \omega ^2}-\sqrt{\frac{p_\varphi^2+\alpha^2}{4}}
-\sqrt{(\frac{p_\varphi}{2}-s)^2+4 \omega ^2},
\label{hcp} \ee where we
introduced  the notation
\be \xi =\frac 1a
\left[\cos\theta-{b}\right],\qquad b = \frac{8 \omega
^2+2s^2-2p_\varphi s- \frac{\alpha^2}{2}}{4h+8\omega ^2+
\frac{\alpha^2}{2}+2s^2},\nonumber
\ee
\be
a=\frac{2}{\sqrt{4h+8\omega^2+\frac{\alpha^2}{2}+2s^2}}
\sqrt{\frac{4h^2-\left(8\omega ^2-p_\varphi s+2s^2\right)
\left(\frac{\alpha^2}{2}+ p_\varphi s\right)}{4h+8\omega
^2+\frac{\alpha^2}{2}+2s^2}-\frac{p_\varphi^2}{2}+p_\varphi s}. \ee
Hence, from (\ref{hcp}) we get that the explicit expression of
 the Hamiltonian has the following dependence from the
action variables: \be
H=\frac{1}{8}\left(2I_2+\sqrt{I_{1}^2+\alpha^2}+\sqrt{(I_1-2s)^2+16\omega^2}\right)^2-\frac{s^2}{2}-
\frac{\alpha^2}{8}-2\omega^2
\ee

The expressions for the angle variables look as follows:
\be
\Phi_1  =
\varphi - \frac{1}{2}\left(\frac{p_\varphi}{\widetilde{p}_\varphi}+
\frac{p_\varphi-2 s}{\sqrt{(p_\varphi-2 s)^2+16 \omega ^2}}\right)
\arcsin \xi +\frac{p_\varphi + s}{\sqrt{(p_\varphi -2 s)^2+16 \omega ^2}}\arctan\eta_+
-\frac{p_\varphi}{\widetilde{p}_\varphi}\arctan\eta_-,
\ee
\be
\Phi_2 =\frac{\partial S}{\partial I_2}  = -\arcsin \xi\;.
\ee
Here we used the notation
 \be
\widetilde{p}_\varphi\equiv\sqrt{p_{\varphi}^2+\alpha^2},\qquad
\eta_\pm\equiv\frac{(1\pm b) \left(\frac{1}{\xi }+
\sqrt{\frac{1}{\xi ^2}-1}\right)\pm a}{\sqrt{(1\pm b)^2-a^2}}
\ee

Finally,  let us restore the radius $r_0$  performing the  replacement (\ref{rep}).
In that case the  Hamiltonian (\ref{alt_magn}) is expressed via action variables as follows:
\be
H=\frac{1}{8 r_0^2}\left(2I_2+\sqrt{I_{1}^2+\alpha^2}+\sqrt{(I_1-2B_0 r^2)^2+
16\omega^2r_0^4}\right)^2-\frac{B_0^2 r_0^2}{2}-\frac{\alpha^2}{8 r_0^2}-2\omega^2r_0^2
\ee
It is seen that, in the planar limit  $r_0\to\infty$,
it results in the Hamiltonian of the Euclidean singular oscillator (\ref{hsaa}).\\

So, we presented the action-angle formulation of the model of the spherical singular oscillator
interacting with a constant magnetic field (\ref{alt_magn}).
The  Hamiltonian of the model is  non-degenerate on
both action variables. But it depends on these variables via
elementary functions in the presence of a constant  magnetic field.

These tell us the area of application of the Higgs oscillator potential and of the $CP^1$ oscillator one.
The Higgs model  is useful for the behavior of the quantum dots systems in the external potential
field, e.g, in the electric field.
The $CP^1$ model should be applied for the study of the behavior of a spherical quantum dots model
in the external magnetic field.

\subsection*{Conclusion}
In this  paper we presented the ``action-angle" formulation of the
textbook two-dimensional singular oscillator model,
 and of its two spherical generalizations, based on the
 so-called Higgs and $CP^1$ spherical oscillator potentials.
 Writing this paper we had in mind two goals:
 the first was the suggestion to use the
 action-angle variables in  the study of quantum dots models.
 Another goal was the suggestion of singular spherical oscillator models
 to the role of confinement potentials in spherical quantum rings.
 In the study of spherical rings we suggested, for the role of the constant magnetic field, the
  magnetic field of the
  Dirac monopole located at the center of the sphere. Surely, the Dirac monopole is a non-physical object.
  However, since we assume to use it for the description of the particles localized on a part of the sphere,
  the non-physical nature of the Dirac monopole can be ignored. The monopole can be considered,
  e.g. as a pole of the magnetic dipole. The possible impact of the Dirac monopole on
  the properties of quantum dots model has been considered, e.g.,
  in \cite{monop}. Besides, magnetic monopoles emerge as a class of magnets known as spin ice \cite{si}.

We restricted ourselves to the formal mathematical formulation of the systems, postponing the consideration of the
particles behavior in the spherical quantum rings, described  by the use of the above presented
models,
for the future study.
\\

{\large Acknowledgments.}
 We are  grateful to Tigran Hakobyan and Haik Sarkisyan   for useful discussions
and comments. The work was supported by
 and ANSEF-2229PS grant and by
 Volkswagen Foundation  grant I/84~496.

\end{document}